\documentstyle[12pt]{article}
\begin{document}
%
\newcommand{\nc}{\newcommand}
\nc{\beq}{\begin{equation}}
\nc{\eeq}{\end{equation}}
\nc{\beqa}{\begin{eqnarray}}
\nc{\eeqa}{\end{eqnarray}}
\nc{\lra}{\leftrightarrow}
\nc{\sss}{\scriptscriptstyle}
{\nc{\lsim}{\mbox{\raisebox{-.6ex}{~$\stackrel{<}{\sim}$~}}}
{\nc{\gsim}{\mbox{\raisebox{-.6ex}{~$\stackrel{>}{\sim}$~}}}
\nc{\rd}{{\rm d}}
\def\G{\Gamma}
\def\etal{{\it et al.}}
\def\l{\lambda}
\def\L{\Lambda}
\def\d{\delta}
\def\tr{\mathop{\rm tr}\nolimits}
\def\pr#1{Phys.~Rev.~{\bf #1}}
\def\np#1{Nucl.~Phys.~{\bf #1}}
\def\pl#1{Phys.~Lett.~{\bf #1}}
\def\prl#1{Phys.~Rev.~Lett.~{\bf #1}}
\def\ie{{\it i.e.}}
\def\Hee{H_{ee}} \def\Hemu{H_{e\mu}} \def\Hmumu{H_{\mu\mu}}
\def\GF{G_{\sss F}}
\def\bra#1{\left\langle #1\right|}
\def\ket#1{\left| #1\right\rangle}
\def\gml{\gamma^\alpha_{\sss L}}
\def\bfH{{\bf H}}
%
\begin{titlepage}
\pagestyle{empty}
\baselineskip=21pt
\rightline{July 1994}
\vskip .4in
\begin{center}
{\large{\bf
Response to Comments on \\
"NEUTRINO FLAVOR EVOLUTION\\
NEAR A SUPERNOVA'S CORE"
}}
\end{center}
\vskip .1in
\begin{center}
Jim Pantaleone\footnote{Permanent address:
Department of Chemistry, Physics and Astronomy,
University of Alaska,
Anchorage, AK  99508}

{\it Physics Department, Indiana University }

{\it Bloomington, IN  47405}

\vskip .2in
\end{center}

\vskip 0.7in
\centerline{ {\bf Abstract} }
\baselineskip=20pt
\vskip 0.5truecm
In a recent paper I concluded that
the neutrino background is important for calculating
neutrino flavor evolution in supernovae, and including it weakens
the connection between r-process nucleosynthesis and cosmologically relevant
neutrino masses.
Comments on this paper have been posted to this
bulletin board; one by J. Cline and the other by Y.Z. Qian and G.M. Fuller.
Here I briefly discuss these Comments and demonstrate that they
are inaccurate and ill-considered.
No corrections to my paper are necessary.

\end{titlepage}
\baselineskip=20pt

In an interesting Letter \cite{PRL}, Qian et. al. used heavy element
nucleosynthesis in supernovae to probe cosmologically significant neutrino
masses.   Therein they described the evolution of neutrino flavors
in a supernovae using standard expressions developed to describe
solar neutrinos (see e.g. \cite{KP}).
However the neutrino background is much larger in
supernovae than in the sun.  In a recent paper \cite{P2} I
included the neutrino background perturbatively, and calculated that
it was not negligible.  It tends to reduce adiabaticity, however reliable
conclusions are difficult because for most of the parameter region
the evolution is very nonlinear.
Hence the connection between neutrino masses and conventional
nucleosynthesis is weakened.
Here I respond to the Comments on my paper which have recently
been posted to this bulletin board.

\section{Response to Cline's Comment.}

In his Comment, Cline \cite{Cline} claims that the off-diagonal elements
in the potential from
forward \(\nu-\nu\) scattering should always be dropped in the
flavor basis.  He argues that {\it flavor} states evolve with definite
phases, and so the off-diagonal terms (which are phase dependent)
will average to zero from summing over many neutrinos.

This arguement is fine for massless neutrinos.
However for massive neutrinos, it is the {\it mass} eigenstates which
evolve with definite phases.  His arguement breaks down then.
To find the neutrino mass eigenstates in a neutrino background is a
nonlinear problem.  Then the flavor off-diagonal elements of the
\(\nu-\nu\) potential must be included and the
summation over neutrinos calculated explicitly.
This is what I proposed in Refs. \cite{P1},
this is what was adopted by subsequent authors
in Refs. \cite{SR}, \cite{Alan}, and even in \cite{Qian},
and it is what is done in my present manuscript.

\section{Response to Qian and Fuller's Comment.}

I will attempt to ignore the hostile tone of Qian and Fuller's  (QF)
\cite{Qian}
Comment, and objectively discuss the three physical points
they raise.

{\bf A)} First, QF claim that in the \(\nu-\nu\) potential
the oscillatory cross terms between
the zeroth order mass eigenstates will always "average to zero".
Thus they maintain that I was "wrong" and "made a conceptual error"
when I included such a term in Eqs. (13) and (14), and then
when I stated in the conclusions that this oscillatory term
could be important.  This criticism by QF is not really
relevant to my paper since I avoided using the oscillatory term
by choosing to do my calculations in the limits where
the amplitude of the oscillations vanish, \(P_c =\) 0 or 1.
However the issue of oscillations is important for future
calculations, so let's examine it.

My expressions for the density matrices
given in Eqs. (13) and (14) are completely correct.
The elimination of arbitrary phases from summation over different
neutrinos at the same point in phase space has
indeed already been carried out.
The phase which remains is that part of the
interference term between the mass eigenstates which is generated
after a neutrino goes through its resonance \cite{PP}.
This phase is the same for all
neutrinos at a point in phase space, so it is not affected
by summation at this stage of the calculation.
QF apparently do not appreciate this point.

When calculating the \(\nu-\nu\) potential at first order in
perturbation theory, one must integrate the zeroth order neutrino density
matrix over phase space.
Then the remaining oscillatory term is certainly greatly reduced in size.
However, when partial adiabaticity occurs,
the oscillatory term only become irrelevant only when
\( < \cos ( \alpha + \beta ) > \) is less than
order \( \sin \theta\)---which ranges as small as \(10^{-3}\).
Hence a very careful,
quantitative calculation of the size of the oscillatory term is
required to establish when it is indeed negligible.
QF give no quantitative estimate to justify neglecting this
oscillatory term.

Furthermore, while neglecting all oscillations
does lead to gross simplification in the description of flavor
evolution, it is not always reasonable.
For example, numerical studies
in the context of the early universe \cite{Alan} have found
oscillatory solutions which are approximately energy independent.
Thus what usually gives the largest reduction in oscillations,
averaging over the energy distribution,
does not work for very nonlinear situations.
This demonstrates that phase effects can not always be neglected.

{\bf B)} In their second point of criticism, QF are extremely vague.

They derive an expression, their Eq. (3), and give it as the
"correct" description of the crossing probability.  However
this expression is identical to the expressions presented in
my paper (in the text at the end of Sect. 2.1, and in the
paragraph containing Eq. (23))!
It is thus difficult to determine
exactly what they are objecting to.

They do stress a difference in how the expression should be
interpreted.   They state that
"The essential nonlinearity of this problem
demands that a self-consistent iteration be performed".
They imply that my method, perturbation theory,
is "wrong" because I have not "iterated" the solution.

However, it is obvious that my method, perturbation theory,
is sufficient for my modest goals of determining
when the problem is nonlinear, and how the nonlinearity tends
to alter the evolution.
My calculations clearly demonstrate that, contrary to the attitude
in Ref. \cite{PRL},  the problem is generally very nonlinear.
Thus QF's present insistence on the necessity of iteration
implicitly supports the results of my paper.
For these reasons, I take QF's second point not as a criticism
but as a grudging endorsement of my paper.

Furthermore, QF's suggested plan of attack, iteration, is not well-defined.
They suggest iterating the density matrix, \(\rho(r,k,\phi)\),
over only a small part of the relevant phase space.
This will lead to a fixed point, but it is generally not unique.
Different iteration procedures will give different physical results.
Thus, when the nonlinearity of the problem is important,
it is difficult to perform reliable calculations.

{\bf C)}As a third criticism, QF object to my statement
that the adiabatic limit
(\(P_c \approx 0\)) is the most appropriate zeroth order solution.
This is because \(P_c \approx 70\%\) for \(E_\nu = 25\)MeV
along the \(Y_e = 0.5 \) line in Fig. 2 of Ref. \cite{PRL}.

However, \(P_c\) is suppressed (approximately exponentially)
and quickly approaches zero i) for values of \(\Delta\)
or \(\sin \theta\) inside the \(Y_e = 0.5\) contour
and ii) for neutrino energies less than 25 MeV.
Suppression (i) applies for all the interesting values of
\(\Delta\) and \(\sin^2 \theta\) except those very near the perimeter
of the  \(Y_e = 0.5 \) contour.   Suppression (ii) applies everywhere
because the \(\nu-\nu\) potential depends on lower energy neutrinos than
nucleosynthesis does.
Hence using the adiabatic limit in the potential is indeed valid.

In summary, QF's Comment is superficial and is completely
irrelevant to my paper.  Their Comment is predominantly a
set of ill-considered suggestions on how to generalize my
work to describe the very nonlinear flavor evolution.
A careful analysis shows that QF now implicitly endorse
most of my results---although they don't acknowledge them.


\end{document}